\documentclass{main}

\usepackage{array}[=2016-10-06]

\usepackage{textcomp}
\usepackage{fancyhdr}
\usepackage{lastpage}
\usepackage{amsmath,amssymb}
\usepackage{cite}

\def\be{\begin{equation}}
\def\ee{\end{equation}}
\def\bea{\begin{eqnarray}}
\def\eea{\end{eqnarray}}

\newcommand{\Photo}{\includegraphics[height=35mm]{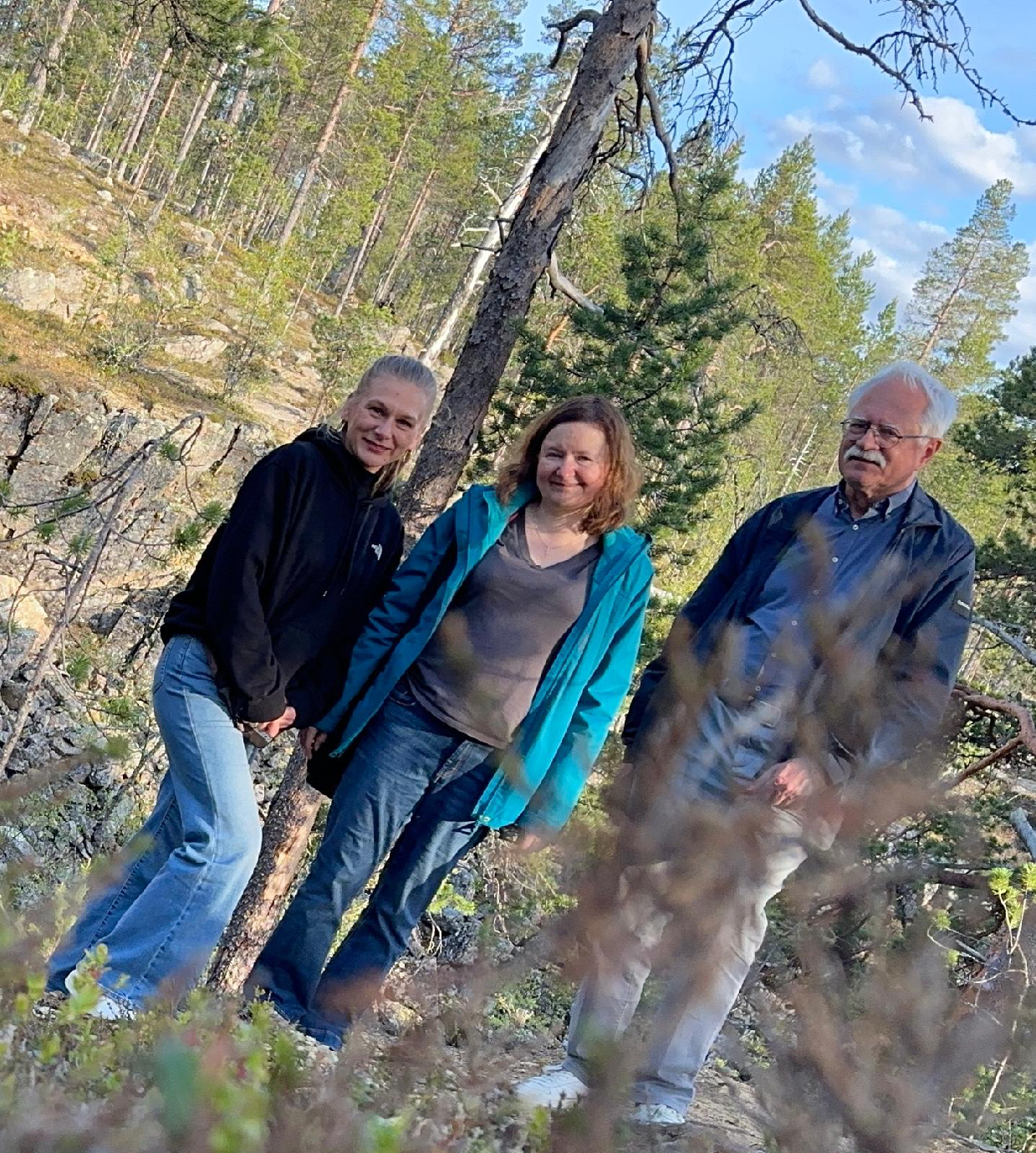}}

\fancypagestyle{firstpagefooter}{%
  \fancyfoot[L]{  \textcopyright \footnotesize This work is an open access article under a Creative Commons Attribution 4.0 International License (https://creativecommons.org/licenses/by/4.0/).}
  \fancyfoot[C]{}  
   
}

\begin{document}

\thispagestyle{firstpagefooter}
\title{\Large From photons to dikaon - theoretical insights into $K^+K^-$ production in nuclear collisions}

\author{\underline{M.~K{\l}usek-Gawenda}\footnote{Speaker, email: mariola.klusek@ifj.edu.pl\\
	Photo taken in the beautiful natural surroundings of the Urho Kekkonen National Park, Saariselkä (Finland). The picture shows the Kraków physics group participating in the conference: Mariola, Iwona Grabowska-Bold and Antoni.}, A.~Szczurek and Nikhil~Krishna}

\address{
Institute of Nuclear Physics PAN, Radzikowskiego 152, Kraków, PL-31-342, Poland
}

\maketitle\abstracts{
The production of kaon pairs can proceed via photoproduction ($\gamma$-Pomeron interaction) or photon–photon fusion ($\gamma\gamma$). An important contribution to this process arises from the decays of scalar, tensor, and vector mesons. This study provides a comprehensive description of dikaon production at both the elementary ($\gamma\gamma \to \mbox{meson} \to K^+K^-$) and nuclear ($Pb Pb \to Pb Pb K^+K^-$) levels. The $\gamma\gamma$ fusion cross section is compared with experimental results from Belle, TPC/Two-Gamma, and ARGUS.
A comparison with existing midrapidity measurements is presented, together with theoretical predictions performed for ultraperipheral Pb–Pb collisions at a nucleon-nucleon center-of-mass energy of $\sqrt{s_{NN}}=5.02$ TeV.}

\footnotesize DOI: \url{https://doi.org/xx.yyyyy/nnnnnnnn}

\keywords{di-kaon production, photon-photon fusion, photoproduction, $\phi$ meson decay}

\section{Introduction}

High-energy photoproduction offers a sensitive probe of strong interaction and the internal structure of hadrons at very small parton momentum fractions $x$.  Ultraperipheral collisions (UPCs) of relativistic heavy ions provide access to photonuclear and two-photon processes.  In such events, the impact parameter between the two nuclei is larger than the sum of their radii, suppressing hadronic overlap and allowing intense electromagnetic fields of the ions that act as beams of quasi-real photons.

Within the vector-meson-dominance (VMD) framework~\cite{Bauer1978,Klein:1999qj}, a photon can fluctuate into a quark-antiquark dipole that scatters elastically from the target nucleus via color-singlet Pomeron exchange, producing a real vector meson with quantum numbers $J^{PC}=1^{--}$.  For light vector mesons such as the $\phi(1020)$, this mechanism leads to exclusive final states composed of charged kaons, $K^+K^-$, originating from the decay $\phi(1020) \rightarrow K^+K^-$ with the branching ratio of $49.1 \%$ \cite{ParticleDataGroup:2022pth}. 

Recent measurements by the ALICE, STAR, LHCb, and CMS Collaborations~\cite{ALICE:2023kgv,STARdata,LHCbdata,CMSdata} 
have provided new insight into exclusive $K^{+}K^{-}$ production in ultraperipheral heavy-ion collisions over a broad invariant mass range.  
Experimental results reveal resonance structures and continuum involvement, but theoretical understanding of the underlying mechanisms remains limited.  
A consistent framework combining photoproduction, dominated by the Drell-S{\"o}ding continuum and the $\phi(1020)$ resonance, and two-photon fusion processes is therefore essential.  
This motivates us to systematically investigate the production of $K^{+}K^{-}$ within the equivalent photon approximation (EPA), with the goal of providing a unified description of the observations at both the elementary and nuclear levels.

\section{Meson photoproduction}

The coherent photoproduction of vector mesons in ultraperipheral heavy-ion collisions (UPCs) is described within the EPA, where the strong electromagnetic fields of relativistic nuclei are treated as fluxes of quasi-real photons~\cite{Klein:1999qj,Baltz:2007kq}. 
In this approach, the total cross section of the $\phi(1020)$ meson in Pb-Pb collisions is written as a superposition of two photon-nucleus interactions, corresponding to photon emission from either of the colliding nuclei:
\begin{equation}
\frac{d\sigma(AA \to AA \phi)}{d^2b dy}
=
\omega_1 N_A(\omega_1,b)\,\sigma(\gamma A \to \phi A;\omega_1)
+
\omega_2 N_A(\omega_2,b)\,\sigma(\gamma A \to \phi A;\omega_2),
\label{eq:AA_dsig_dy}
\end{equation}
where $M_{\phi}$ is the $\phi(1020)$ mass, $y$ is the meson rapidity, and $N_A(\omega_i,b)$ denotes the photon flux with the photon energy $\omega_i$ associated with the electromagnetic field of the nucleus $A$.  
The function $\sigma(\gamma A \to \phi A;\omega_i)$ represents the photon-nucleus cross section, which is obtained by integrating the differential distribution over the momentum transfer $t$:
\begin{equation}
\sigma(\gamma A \to \phi A)
=
\int dt\,
\frac{d\sigma(\gamma A \to \phi A)}{dt}.
\label{eq:sigma_gammaA}
\end{equation}

The photon-nucleus process is connected to the elementary photon-nucleon reaction through the vector-meson dominance (VMD) model and Glauber multiple-scattering formalism. 
The differential cross section for $\gamma N \to \phi N$ is related to the total $\phi N$ cross section via the optical theorem:
\begin{equation}
\left.\frac{d\sigma(\gamma N \to \phi N)}{dt}\right|_{t=0}
=
\frac{\alpha_{\mathrm{em}}}{4\,f_{\phi}^{2}}\,(1+\eta^{2})\,
\sigma_{\mathrm{tot}}^{2}(\phi N),
\label{eq:gammaN_forward}
\end{equation}
where $\alpha_{\mathrm{em}}$ is the electromagnetic coupling constant, $f_{\phi}$ is the $\phi$-photon coupling parameter determined from the $\phi \to e^{+}e^{-}$ decay width, and $\eta$ is the ratio of real to imaginary parts of the forward scattering amplitude.  
The energy dependence of $\sigma_{\mathrm{tot}}(\phi N)$ can be parametrized following Regge-type fits to $\gamma p$ data~\cite{Klein:1999qj}.

The nuclear cross section is obtained by folding the elementary amplitude with the elastic nuclear form factor, including nuclear shadowing effects. 
The corresponding total $\phi$-nucleus cross section is given by
\begin{equation}
\sigma_{\mathrm{tot}}(\phi A)
=
\int d^{2}\mathbf{b}\,
\left[
1-
\exp \left(-
\sigma_{\mathrm{tot}}(\phi N)\,T_A(\mathbf{b})\right)
\right],
\label{eq:sigtot_phiA}
\end{equation}
where $T_A(\mathbf{b})=\int dz\,\rho_A(\sqrt{b^{2}+z^{2}})$ is the nuclear thickness function normalized to $\int d^{2}\mathbf{b}\,T_A(\mathbf{b})=A$, and $\rho_A(r)$ is the nuclear density profile (typically Woods-Saxon form).  

The total photon-nucleus cross section can then be obtained by integrating Eq.~(\ref{eq:sigma_gammaA}) over $t$, which is limited by the coherence condition $|t| \lesssim 1/R_A^{2}$:
\begin{equation}
\sigma(\gamma A \to \phi A)
=
\left.\frac{d\sigma(\gamma N \to \phi N)}{dt}\right|_{t=0}
\int dt\,|F_A(t)|^{2}\,,
\label{eq:gammaA_sigma_int}
\end{equation}
where $F_A(t)$ is the elastic nuclear form factor.
Finally, this formalism, combining the VMD, Glauber, and EPA approaches, provides the basis of the calculations. 

The invariant mass distribution of meson pairs produced in photon-induced reactions is governed by the interplay between resonant and nonresonant processes. 
In the simplest case, the formation of a vector meson such as the $\phi(1020)$ is described by a relativistic Breit-Wigner function, 
which represents the propagation of an unstable particle with mass $M_\phi$ and a mass-dependent width $\Gamma_\phi(M_{KK})$. 
This parametrization yields a pure Breit–Wigner line shape centered at the pole mass of the resonance and provides an excellent description when the production mechanism is dominated by a single resonant amplitude.
In practice, however, the same final state can also be produced directly, without the formation of an intermediate resonance. 
In the case of the $K^+K^-$ system, the photon may couple to the kaon pair via a nonresonant continuum amplitude, 
which interferes with the resonant $\phi(1020)$ contribution. 
This interference is described by the Drell-S{\"o}ding formalism~\cite{Drell:1966cu,Soding:1965nh}, 
in which the total amplitude is expressed as the coherent sum of the two components, 
\begin{equation}
\mathcal{A}(M_{KK}) = \mathcal{A}_{\phi}(M_{KK}) + \mathcal{A}_{\mathrm{cont}}(M_{KK})\,e^{i\varphi}.
\label{eq:ds_amp}
\end{equation}
Here, $\mathcal{A}_{\phi}$ denotes the Breit-Wigner amplitude associated with the $\phi(1020)$ resonance, 
$\mathcal{A}_{\mathrm{cont}}$ represents the direct $K^+K^-$ continuum, and $\varphi$ is the relative phase between them.
The corresponding cross section is proportional to the squared modulus of this sum, 
leading to three distinct contributions: a resonant term, a smooth continuum, and an interference term.
The interference modifies the apparent shape of the resonance, producing an asymmetric distortion of the line and a shift of the observed peak position with respect to the true pole mass. 
This phenomenon, known as the Drell-S{\"o}ding effect, is most prominent when the magnitudes of the resonant and nonresonant amplitudes are comparable.

The pure Breit-Wigner case is recovered in the limit of a vanishing continuum amplitude, $\mathcal{A}_{\mathrm{cont}} \to 0$, 
which eliminates the interference and restores symmetric Breit-Wigner profile. 
When the continuum term is finite, the relative phase $\varphi$ determines the character of the interference: 
a constructive phase ($\varphi \approx 0$) enhances the high-mass tail of the distribution, 
whereas a destructive phase ($\varphi \approx \pi$) suppresses it. 
Such effects have been observed in photoproduction and two-photon production of meson pairs~\cite{ALEPH2003,Uehara2013}, 
including $\rho^0 \rightarrow \pi^+\pi^-$ and $\phi(1020) \rightarrow K^+K^-$ decays.

The resonant component is represented by a relativistic Breit-Wigner distribution in the Jackson form \cite{Jackson1964},
\begin{equation}
\mathcal{A}_{\phi}(M_{KK}) =
A_{\phi}\,
\frac{\sqrt{M_{KK}\,M_{\phi}\,\Gamma_{\phi}(M_{KK})}}
{M_{KK}^2 - M_{\phi}^2 + i\,M_{\phi}\,\Gamma_{\phi}(M_{KK})},
\label{eq:DS_breitwigner}
\end{equation}
where $A_{\phi}$ contains the production dynamics and the nuclear form factor. 
The mass-dependent width for a vector meson decaying into two pseudoscalars (P-wave) is given by
\begin{equation}
\Gamma_{\phi}(M_{KK}) =
\Gamma_{\phi}^0
\left(\frac{q(M_{KK})}{q(M_{\phi})}\right)^3
\left(\frac{M_{\phi}}{M_{KK}}\right),
\qquad
q(M)=\frac{1}{2}\sqrt{M^2 - 4m_{K}^2}.
\label{eq:DS_width}
\end{equation}

In the photoproduction of the $\phi(1020)$ meson, as observed in the recent ALICE measurements~\cite{ALICE:2023kgv},  
the invariant mass distribution of the $K^{+}K^{-}$ system cannot be reproduced by a pure Breit-Wigner resonance alone.  
A realistic description requires accounting for the distortion of the resonance profile caused by the interference between the resonant $\phi(1020)$ amplitude and the nonresonant continuum of direct charged kaon pair production.  
This effect introduces an effective mass smearing that modifies both the apparent width and the symmetry of the resonance peak.  
Only when this interference-induced broadening is included does the theoretical line shape match the experimentally measured $K^{+}K^{-}$ spectrum, see Fig.~\ref{fig:dsig_dmkk}.

\section{Two-photon fusion}

A solid theoretical foundation for the description of exclusive meson-pair production in 
two-photon fusion has been established in the context of the $\gamma\gamma \to \pi^{+}\pi^{-}$ 
reaction. In Ref.~\cite{Klusek-Gawenda:2013rtu}, the energy and angular dependence of the dipion system was successfully modeled over a wide energy range by combining 
the Born continuum amplitudes (contact, $t$- and $u$-channel exchanges) with 
$s$-channel resonance contributions, including nine resonances. The approach further incorporates hadronic form factors that suppress the 
point-like QED contributions at higher energies, preserving gauge invariance and reflecting 
the composite structure of the mesons. 
When applied to the elementary process $\gamma\gamma \to \pi^{+}\pi^{-}$,  
this formalism provides an excellent quantitative description of the experimental cross sections and angular distributions obtained by the Belle, ALEPH, and other collaborations over a wide energy range, extending up to about $W_{\gamma\gamma} \simeq 3~\text{GeV}$.  
The remarkable agreement between the theoretical predictions and the measured data confirms the robustness of the approach and its ability to capture both the resonant structures and the smooth continuum contribution.  
Such consistency between model and experiment demonstrates that the same methodology can be confidently applied to the $\gamma\gamma \to K^{+}K^{-}$ reaction,  
where analogous mechanisms - Born terms, resonance contributions, hand-bag model \cite{Diehl:2010} and interference effects are expected to govern the dynamics of strange-meson production.
In the case of kaons, the higher production threshold and the involvement of strange quarks lead to a richer resonance structure and a more complex interplay between the continuum and resonant contributions. 

It should be noted that the present understanding of the resonance spectrum contributing to the $\gamma\gamma \to K^{+}K^{-}$ reaction remains rather limited.  
In particular, the experimental knowledge of the two-photon decay widths and branching ratios into $K^{+}K^{-}$ channels is still incomplete and often affected by large uncertainties.  
Consequently, the choice of resonances to be included in the phenomenological description cannot be fully constrained by existing data.  
In this study, we focus on a representative subset of states: $f_{2}(1525)$, $a_{2}(1750)$, $f_{2}(2010)$, $f_{2}(2300)$, which are known to couple to kaon pairs and are expected to provide the dominant contributions in the invariant mass region up to about $2.5~\text{GeV}$.  
These resonances have been identified and characterized in the measurements of the Belle Collaboration~\cite{Belle},  
and they capture the essential features of the strange-meson sector while minimizing the model dependence introduced by poorly known higher-mass states.

\begin{figure}[!h]
	\centering
	\includegraphics[scale=0.35]{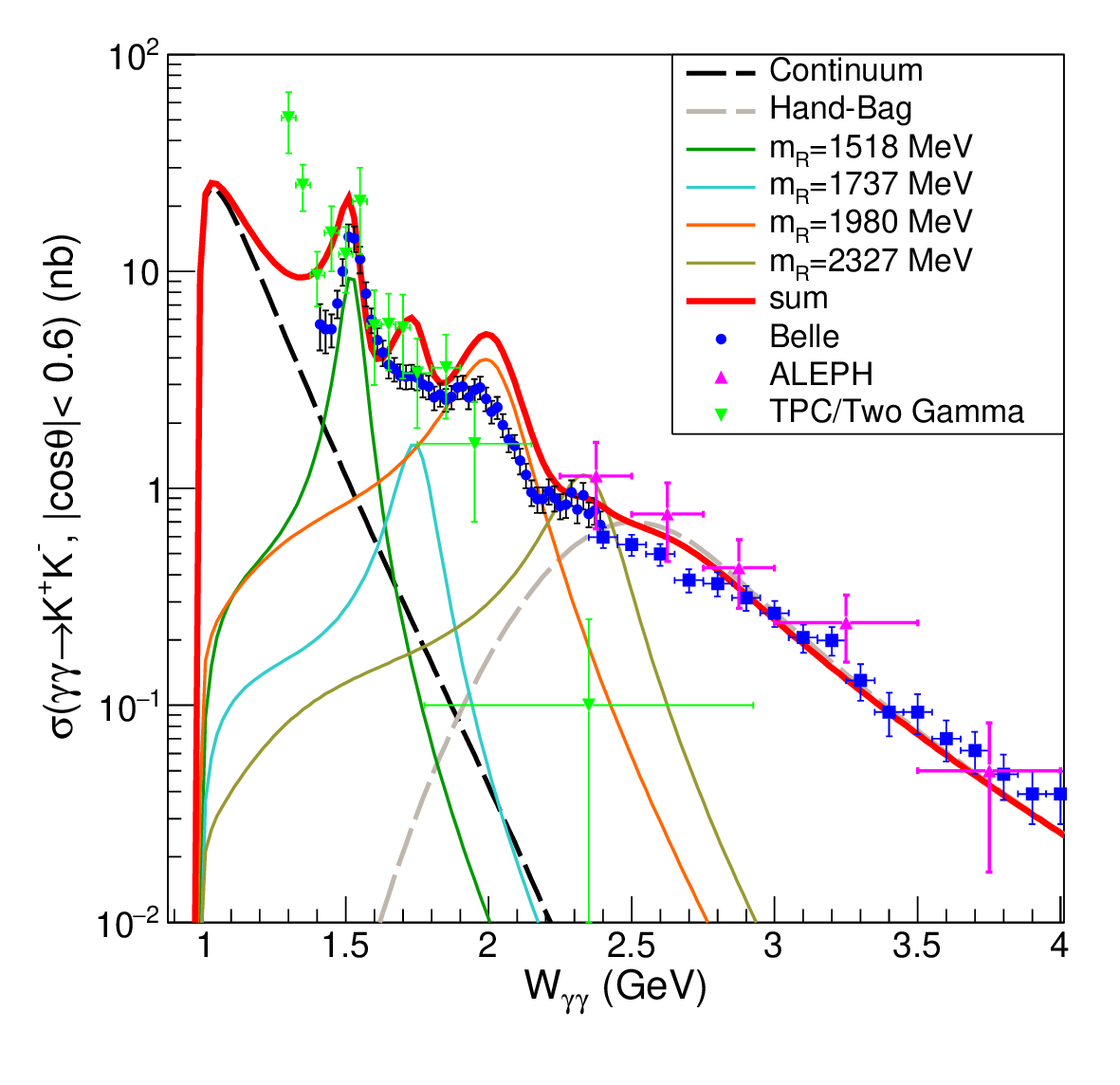}
	\caption
	{Differential cross section for the $\gamma\gamma \to K^{+}K^{-}$ process as a function of the two-photon center-of-mass energy $W_{\gamma\gamma}$ for $|\cos\theta| < 0.6$.  
			Theoretical curves correspond to individual resonance contributions from $f_{2}(1525)$, $a_{2}(1750)$, $f_{2}(2010)$, and $f_{2}(2300)$,  
			together with the nonresonant continuum and the contribution from the hand-bag mechanism. The solid red line represents the coherent sum of all components. 
			Experimental data points from the Belle~\cite{Belle,Belle2}, ALEPH~\cite{ALEPH2003}, and TPC/Two-Gamma experiments are shown for comparison. }  
		\label{fig:el_xces_06}
\end{figure}
 
Near threshold, the cross section is dominated by the vector $\phi(1020)$ meson, which decays predominantly into $K^{+}K^{-}$, see Fig. \ref{fig:el_xces_06}.  
At higher invariant masses, the spectrum exhibits several resonances contributing through $s$-channel formation:  
$f_{2}(1525)$, $a_{2}(1750)$, $f_{2}(2010)$, $f_{2}(2300)$.
These states collectively shape the invariant mass dependence of the $\gamma\gamma \to K^{+}K^{-}$ cross section.  
A coherent treatment of these contributions, together with the nonresonant Born terms and hand-bag model (dashed lines), ensures a realistic description of both the total and differential cross sections over the energy range up to about $W_{\gamma\gamma}=4~\text{GeV}$.
The same formalism can then be embedded in the EPA to describe exclusive $K^{+}K^{-}$ production in ultraperipheral heavy-ion collisions,  
linking the elementary $\gamma\gamma$ dynamics with the observables measured in experiments such as STAR, ALICE, LHCb, and CMS.

The total cross section for exclusive di-kaon production in ultraperipheral heavy-ion collisions 
can be formulated within the EPA directly in the impact-parameter representation.  
In this approach, the electromagnetic fields of the colliding nuclei are treated as fluxes of quasi-real photons distributed in the transverse plane.  
The two photons, emitted coherently may interact to produce a $K^{+}K^{-}$ pairs. Nuclei that remain separated by more than twice their radii avoid hadronic overlap.
The corresponding total cross section is given by
\begin{equation}
\sigma_{AA\to AA\,K^{+}K^{-}}
=
\int d^{2}\mathbf{b}_1\,d^{2}\mathbf{b}_2\;
d\omega_1\,d\omega_2\;
\Theta\!\big(|\mathbf{b}_1-\mathbf{b}_2|-2R_A\big)\,
N(\omega_1,\mathbf{b}_1)\,N(\omega_2,\mathbf{b}_2)\;
\sigma_{\gamma\gamma\to K^{+}K^{-}}(W_{\gamma\gamma}),
\label{eq:EPA_basic}
\end{equation}
where $N(\omega_i,\mathbf{b}_i)$ denotes the photon flux generated by nucleus $A$ at a distance $\mathbf{b}_i$,  
and $\omega_i$ are the photon energies related to the two-photon invariant mass and rapidity.
The step function $\Theta(|\mathbf{b}_1-\mathbf{b}_2|-2R_A)$ ensures that the interaction is purely electromagnetic by excluding configurations with overlapping nuclei.  
The elementary cross section $\sigma_{\gamma\gamma\to K^{+}K^{-}}(W_{\gamma\gamma})$  
is obtained from the coherent sum of the Born continuum, resonance contributions and their interference terms, as discussed in the previous section.  
This formulation provides the most direct connection between the elementary two-photon dynamics and the experimentally measured cross sections in Pb-Pb UPC.

\section{Conclusion}

\begin{figure}[!h]
	\centering
	\includegraphics[scale=0.35]{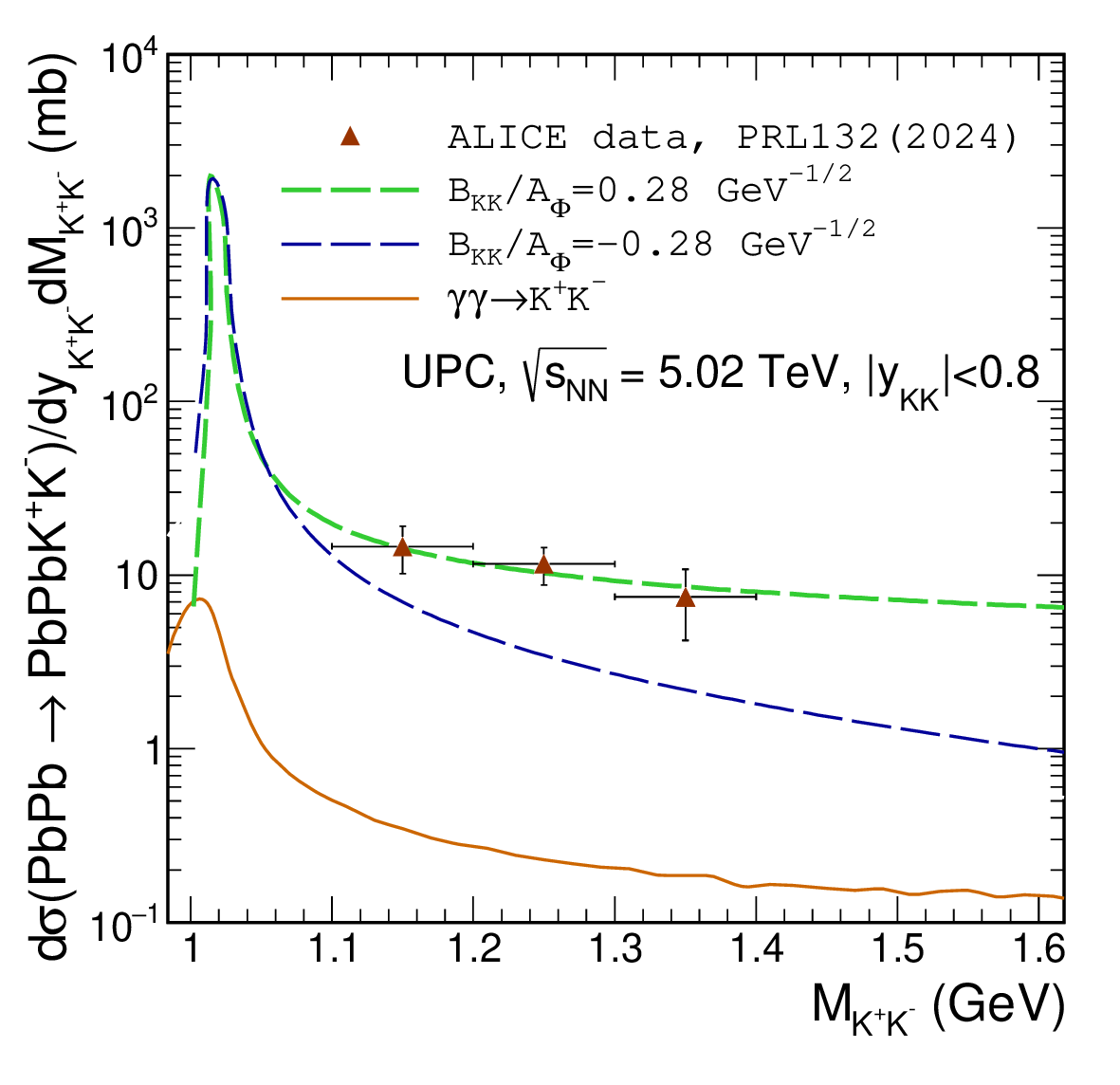}
	\caption{Differential cross section as a function of di-kaon invariant mass for UPC of $Pb-Pb$ at the energy of $\sqrt{s_{NN}}=5.02$~TeV.}
	\label{fig:dsig_dmkk}
\end{figure}

A key result of this study is the nuclear cross section for $Pb Pb \to PbPbK^+K^-$, 
which simultaneously illustrates the contributions from photoproduction and from two-photon fusion processes.  Figure~\ref{fig:dsig_dmkk} shows the differential cross section 
$d\sigma/dM_{K^+K^-}$ for exclusive $K^+K^-$ photoproduction in ultraperipheral Pb-Pb collisions at $\sqrt{s_{NN}}=5.02$~TeV. 
A clear peak corresponding to the $\phi(1020)$ resonance is observed above a smooth continuum extending over the full invariant mass range. 
The shape of the spectrum is well described by the coherent sum of a resonant Breit-Wigner function and a nonresonant continuum amplitude according to the Drell-S{\"o}ding formalism. The curves correspond to two different relative phases between the resonant and continuum amplitudes, 
$|B/A| = \pm 0.28~\text{GeV}^{-1/2}$, which affect the interference pattern and the line shape near the $\phi(1020)$ resonance.  
Experimental data points from the ALICE Collaboration~\cite{ALICE:2023kgv} at $M_{KK}=1.1-1.4~\text{GeV}$ are shown for comparison.   
The interference between these two components leads to a characteristic asymmetry of the $\phi(1020)$ line shape and a small shift of the apparent peak position with respect to the nominal mass value reported by the PDG~\cite{ParticleDataGroup:2022pth}.

Recently, the STAR~\cite{STARdata}, LHCb~\cite{LHCbdata}, and CMS~\cite{CMSdata} collaborations have shown new measurements of kaon-pair production 
in ultraperipheral collisions across a broad invariant mass spectrum, $M_{KK}=(1-5)~\text{GeV}$, 
revealing distinct peaks corresponding to higher-mass resonances.  
Since the parameters of many of these states are not well constrained in the PDG, 
a refined theoretical interpretation is crucial.  
The key observable, the invariant mass distribution $d\sigma/dM_{K^{+}K^{-}}$, 
compares the interplay between photoproduction and photon-fusion dynamics, 
and its simultaneous description offers a unified view of exclusive $K^{+}K^{-}$ production from the elementary to the nuclear collisions.

In addition, a complete description of the $\gamma\gamma \to K^{+}K^{-}$ process should ideally account not only for the energy spectrum but also for the angular distributions measured at different two-photon energies.  
Such an analysis, however, is far from trivial, as interference effects between the continuum and multiple overlapping resonances play a decisive role in shaping the observed angular dependence.  
Despite its complexity, performing this combined fit in both energy and scattering angle remains a desirable goal,  
since it would provide a comprehensive characterization of the elementary $\gamma\gamma \to K^{+}K^{-}$ reaction and thus deliver the necessary input for modeling the corresponding process at the nuclear level in UPC.  
Recent preliminary results from the LHCb Collaboration \cite{LHCbdata}  
indicate that an accurate description of the data requires the inclusion of a broader spectrum of resonances than previously considered.

\section*{Acknowledgments}

This work is supported in part by the National Science Center (NCN), Poland, SONATA BIS grant no. 2021/42/E/ST2/00350.

\end{document}